# Observation of the gradual transition from one-dimensional to two-dimensional Anderson localization

U. Naether,[1,2,6] Y. V. Kartashov,[3,4] V. A. Vysloukh,[5] S. Nolte,[6] A. Tünnermann,[6] L. Torner,[3] and A. Szameit[6]

[1]Departamento de Física, Faculdad de Ciencias, Universidad de Chile, Santiago, Chile
[2]Center for Optics and Photonics (CEFOP), Casilla 4016, Concepción, Chile
[3]ICFO-Institut de Ciencies Fotoniques, and Universitat Politecnica de Catalunya, Mediterranean Technology Park, 08860 Castelldefels (Barcelona), Spain
[4]Institute of Spectroscopy, Russian Academy of Sciences, Troitsk, Moscow Region, 142190, Russia
[5]Departamento de Fisica y Matematicas, Universidad de las Americas – Puebla, 72820, Puebla, Mexico
[6]Institute of Applied Physics, Friedrich-Schiller-Universität Jena, Max-Wien-Platz 1, 07743 Jena, Germany

We study the gradual transition from one-dimensional to two-dimensional Anderson localization upon transformation of the dimensionality of disordered waveguide arrays. An effective transition from one- to two-dimensional system is achieved by increasing the number of rows forming the arrays. We observe that, for a given disorder level, Anderson localization becomes weaker with increasing number of rows, hence the effective dimension.

Unique physical phenomena attributed to the reduced dimensionality are encountered where a global wave function is confined in a lower-dimensional system. They appear in various physical contexts, such as quantum wells, quantum wires, quantum dots, and periodic or quasi-periodic semiconductor superlattices, to name a few [1-3]. One of such phenomena that strongly depends on the dimensionality of the system is Anderson localization [4], a topic that generates a continuously renewed interest. Although Anderson localization has been observed in various systems [5-9], optical waveguide arrays set an excellent model system where it has been observed in two-dimensional (2D) [10] and one-dimensional (1D) settings [11-13]. However, not much is known so far about the transition from 1D to 2D dimensionalities. To the best of our knowledge, only in a visionary theoretical work published almost three decades ago the transition of the effective width of a wave function from a 1D setting to a 2D setting was analyzed [14]. In this Letter, we report the first experimental observations which are consistent with the predictions of [14] and thus study how the confinement of the wave function in one direction impacts Anderson localization. In our experiment we employed arrays of evanescently coupled waveguides where a varying number of waveguide layers simulates the transition from 1D to 2D dimensionality [15]. Our results show generic features, thus similar effects should appear in magnetic nonlinear chains, molecular crystals, nonlinear meta-materials, or matter-waves held in optical lattices with variable effective dimensionality.

To analyze the results of the experiment, we model numerically light propagation in disordered waveguide arrays governed by the Schrödinger equation for the dimensionless light field amplitude $q$:

$$i\frac{\partial q}{\partial \xi} = -\frac{1}{2}\left(\frac{\partial^2 q}{\partial \eta^2} + \frac{\partial^2 q}{\partial \zeta^2}\right) - pR(\eta,\zeta)q, \qquad (1)$$

where $\eta, \zeta$ are the transverse coordinates normalized to the characteristic beam width and $\xi$ is the propagation distance normalized to the diffraction length. The parameter $p$ describes the refractive index contrast. The refractive index distribution in the waveguide array is described by the function $R(\eta,\zeta) = \sum_{k,m} \exp[-(\eta-\eta_k)^4/g_\eta^4 - (\zeta-\zeta_m)^4/g_\zeta^4]$, where $g_\eta, g_\zeta$ are the widths of waveguides along the horizontal $\eta$ and vertical $\zeta$ axes, respectively. In our simulations we assume, that the array is infinite in the horizontal direction and has a finite number of rows in the vertical direction, i.e. in the absence of disorder the waveguide positions are given by $\eta_k = kd_\eta$ and $\zeta_m = md_\zeta$. Here $d_\eta, d_\zeta$ stand for regular waveguide spacing along the $\eta$ and $\zeta$ axes, $k \in \mathbb{Z}$, while $m \in [-n/2, +n/2]$ for even number of rows and $m \in [-(n-1)/2, +(n-1)/2]$ for odd number of rows $n$. Importantly, the effective dimensionality of such an array can be controlled by increasing the number of rows: for $n=1$ one gets 1D array, for moderate $n$ values the array's dimensionality can be considered as intermediate between 1 and 2, while the transition to a truly 2D array occurs at $n \to \infty$. Note that this approach is different from a recent theoretical analysis, where a strong anisotropy of Anderson localization was addressed [16]. We introduce disorder into the spacing between the waveguides. In disordered arrays the waveguide positions are given by $\eta_k = kd_\eta + \delta\eta_{km}$, $\zeta_m = md_\zeta + \delta\zeta_{km}$, where the uncorrelated random shifts $\delta\eta_{km}, \delta\zeta_{km}$ of the waveguide centers along the $\eta$ and $\zeta$ axes are uniformly distributed within the segments $[-S_\eta, +S_\eta]$ and $[-S_\zeta, +S_\zeta]$, respectively. The disorder level in our system is therefore controlled by the parameters $S_\eta < d_\eta/2$ and $S_\zeta < d_\zeta/2$; they are limited in order to avoid overlap between neighboring waveguides.

The parameters of the arrays in Eq. (1) were selected in accordance to the experiment. Thus, the waveguide widths are $g_\eta = 0.3$ and $g_\zeta = 0.53$ (corresponding to $3 \times 5.3$ $\mu m^2$ wide elliptical waveguides, due to the fabrication procedure). The spacing $d_\eta = 1.5$ and $d_\zeta = 1.7$ corresponds to the actual horizontal and vertical waveguide separations of 15 $\mu m$ and 17 $\mu m$, respectively (such a spacing yields almost equal horizontal and vertical rates of discrete diffraction in regular two-dimensional arrays). The refractive index contrast $p = 18$ corresponds to an actual refractive index modulation of $\sim 1.2\times 10^{-3}$ at the wavelength $\lambda = 800$ nm and the 100 mm length of our experimental samples corresponds to the propagation distance $L = 87.2$. We fixed the level of disorder at $S_\eta = 0.5$ and $S_\zeta = 0.3$ (that limits maximal possible random shift of waveguides to 5 $\mu$m in horizontal and to 3 $\mu$m in vertical directions and provides a comparable localization degree along $\eta$ and $\zeta$ axes in two-dimensional arrays). We will be interested in the impact of the array dimensionality on Anderson localization.

In order to illustrate Anderson localization in numerical simulations we utilize a Monte-Carlo approach and compute $Q = 10^3$ realizations of disordered arrays for each number of rows $n$ between 1 and 20. For each realization and amount of layers Eq. (1) was solved with the input conditions $q(\eta,\zeta,\xi = 0) = w(\eta,\zeta)$, where the function $w(\eta,\zeta)$ describes a linear guided mode of an isolated waveguide that is located closest to the center of array. For each $n$ value we calculated the intensity distribution $I_{av}(\eta,\zeta,\xi)$, its horizontal integral width $w_\eta(\xi)$, and the form-factor $\chi(\xi)$. All these quantities are averaged over the ensemble of all array realizations:

$$I_{av} = Q^{-1}\sum_{i=1,Q}|q_i|^2,$$
$$w_\eta^2 = U^{-1}Q^{-1}\sum_{i=1,Q}\iint \eta^2 |q_i|^2 d\eta d\zeta, \quad (2)$$
$$\chi^2 = U^{-2}Q^{-1}\sum_{i=1,Q}\iint |q_i|^4 d\eta d\zeta,$$

where $U = \iint |q_i|^2 d\eta d\zeta$ is the total conserved energy flow. Please note, that while parameter $w_\eta(\xi)$ characterizes the evolution of the width of the distribution along the $\eta$ axis taking into account also small-amplitude radiation moving toward the periphery of the array, the inverse form-factor $\chi^{-1}(\xi)$ characterizes the total width of most intensive "localized" fraction of $|q_i(\eta,\zeta,\xi)|^2$ distribution, disregarding small-amplitude radiation.

The averaged intensity distributions for waveguide arrays with $n = 1, 3, 5$ and 17 rows at the sample output ($\xi = L$) are shown in Fig. 1. The localization around the excited waveguide is apparent in all cases - for the same sample length without disorder one would observe considerable discrete diffraction with main lobes of diffraction pattern located around waveguides with $k,m \sim 8$. The inspection of output intensity distributions reveals, that they are exponentially localized, a signature of Anderson localization. The averaged output intensity distribution notably expands in vertical direction with the increase of the number of rows in the array. This expansion is most pronounced for moderate number of rows up to the value $n \sim 9$, above which the output patterns become practically indistinguishable for increasing $n$. Surprisingly, the horizontal width $w_\eta$ of the pattern also increases with $n$ although the array expands only in vertical direction. This suggests that the dimensionality of the array in one direction does affect localization in the orthogonal direction. One can clearly observe this in Fig. 2, showing evolution of the inverse form-factor and $\eta$-width with propagation distance and illustrating the transition from initial ballistic spreading to localization. Since $\chi^{-1}$ characterizes the width of the most intense fraction of $|q_i|^2$ distribution, which arises due to the excitation of Anderson modes in the direct proximity of the input waveguide, the transition to localization is most apparent from $\chi^{-1}(\xi)$ de-pendence [Fig. 2(a)]. In arrays featuring more rows, the beam expansion is faster and larger distances are required to achieve a regime with suppressed light transport across the array [Fig. 2(b)].

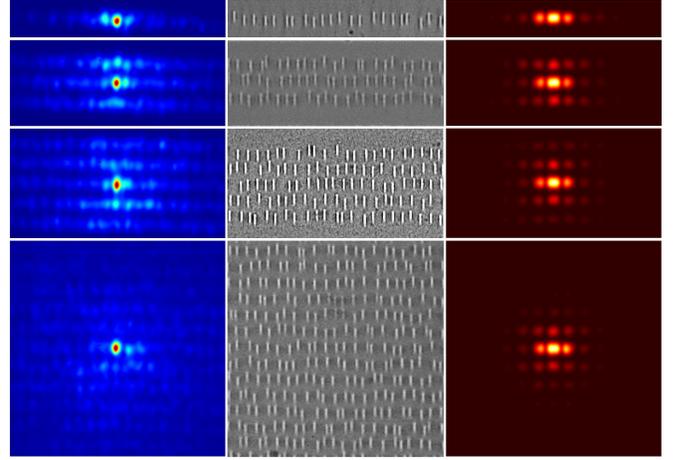

Fig. 1. Experimentally measured averaged output intensity distributions (left), microscopic images of waveguides arrays (center), and theoretically calculated averaged output intensity distributions (right) in disordered waveguide arrays with 1, 3, 5, and 17 rows (from top to bottom).

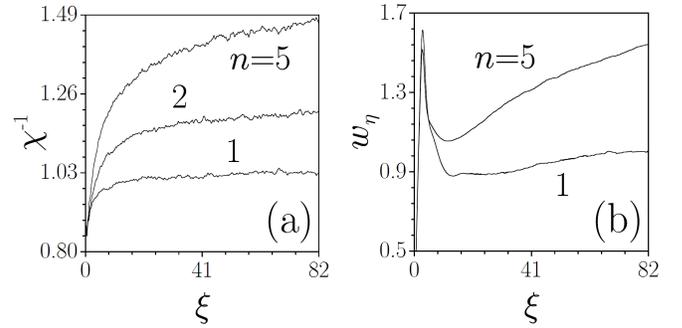

Fig. 2. Dependencies of inverse form-factor (a) and width along the $\eta$-axis (b) averaged over $10^3$ array realizations on propagation distance at $S_\eta = 0.5$, $S_\zeta = 0.3$.

The central prediction of the numerical studies is depicted in Fig. 3(a), which illustrates the dependence of the inverse averaged form-factor (the width of localized beam fraction) at the output face of the sample on the number of rows $n$ (effective dimensionality) of the array. Starting from a small $\chi^{-1}$ value for the one-dimensional case, it grows rapidly with $n$, but saturates around the value $n \sim 9$, above which the system can be considered effectively two-dimensional. Thus, the localization in one-dimensional case is always stronger than in two-dimensional one. It should be stressed that this dependence is obtained for a distance corresponding to the length of our experimental samples, but it remains practically unchanged for much larger $\xi$ values. This result confirms the prediction of [14] where localization strength was also found to decrease with increase of the dimensionality of the system. Notice however the difference to the results in [16], where somewhat different system with strongly anisotropic coupling in two orthogonal directions was used and disorder was introduced in both lattice period and depths of lattice sites.

We fabricated waveguide arrays by the laser direct-writing technology [17] in fused-silica glass. Using a 800 nm wavelength femto-second laser focused around 250 microns below the glass surface, permanent refractive index changes were induced resulting in the formation of elliptically shaped waveguides. We prepared six disordered arrays of $n \times 81$ waveguides, with $n = [1,3,5,7,11,17]$ layers and a total length of 100 mm. The microscopic images of our arrays are shown in the central column of Fig. (1). The mean distance in the horizontal (vertical) direction was 15 $\mu$m (17 $\mu$m). The disorder was introduced by a distance variation of $\delta \times 5$ $\mu$m ($\delta \times 3$ $\mu$m), respectively, with $\delta$ being the random number uniformly distributed in the interval $[-1,+1]$. In each array 30 different individual waveguides located far from the borders were excited using a Ti: Sapphire laser system with 800 nm wavelength at low input power to ensure linear propagation. At the end facet, the intensity patterns were recorded with a CCD camera and averaged over each of the 30 realizations.

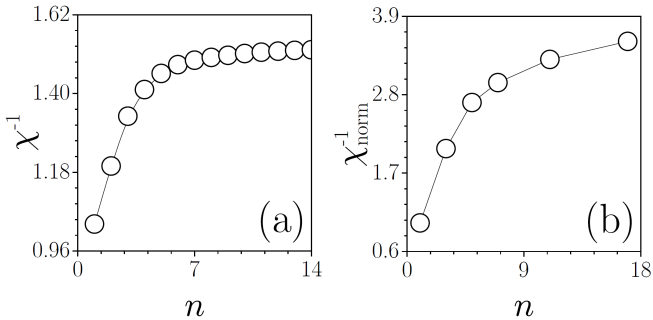

Fig. 3. Theoretically calculated (a) and experimentally measured (b) output averaged inverse form-factor versus number of rows at $S_\eta = 0.5$, $S_\zeta = 0.3$. The inverse form-factor in (b) is normalized in such way that $\chi^{-1}_{norm} = 1$ at $n = 1$.

Examples of the experimental averaged output intensity distributions are shown in the left column of Fig. 1 for the arrays with 1,3,5, and 17 rows. Except for more pronounced background the experimental images demonstrate close similarities with theoretically calculated distributions shown in the last column of the same figure. Furthermore, for each experimental intensity pattern the form-factor was computed and averaged following Eq. (2). The resulting dependence $\chi^{-1}(n)$ is shown in Fig. 3(b). In complete agreement with the results of theoretical simulations we observe rapid initial growth of the width of output intensity distribution with increase of $n$ and the tendency for its saturation for large values of $n \sim 10$.

In conclusion, we have studied for the first time the gradual transition from 1D to 2D Anderson localization by using an effective intermediate dimensionality that is obtained by a varying the amount of waveguide layers. Our observations indicate that Anderson localization of light becomes weaker with the increase of the effective dimensionality, a fundamental result that is relevant to all areas of science where Anderson localization occurs.

The authors appreciate the financial support of a CONICYT doctoral fellowship and acknowledge financial support from the German Ministry of Education and Research (Center for Innovation Competence program, grant 03Z1HN31).

**References**
1. W. E. Buhro and V. L. Colvin, Nature Mater. **2**, 138 (2003).
2. L. Esaki and R. Tsu, IBM J. Res. Dev. **14**, 61 (1970).
3. H. Rodriguez-Coppola, Microelectron. J. **33**, 379 (2002).
4. P. W. Anderson, Phys. Rev. **109**, 1492–1505 (1958).
5. D. S. Wiersma, P. Bartolini, A. Lagendijk, and R. Righini, Nature **390**, 671 (1997).
6. J. Billy, V. Josse, Z. Zuo, A. Bernard, B. Hambrecht, P. Lugan, D. Clement, L. Sanchez-Palencia, P. Bouyer, and A. Aspect, Nature **453**, 891 (2008).
7. G. Roati, C. D'Errico, L. Fallani, M. Fattori, C. Fort, M. Zaccanti, G. Modugno, M. Modugno, and M. In guscio, Nature **453**, 895 (2008).
8. H. Hui, A. Strybulevych, J. H. Page, S. E. Skipetrov, and B. A. van Tiggelen, Nature Phys. **4**, 945 (2008).
9. J. Chabe, G. Lemarie, B. Gremaud, D. Delande, P. Szriftgiser, and J. C. Garreau, Phys. Rev. Lett. **101**, 255702 (2008).
10. T. Schwartz, G. Bartal, S. Fishman and M. Segev, Nature **466**, 52 (2007).
11. Y. Lahini, A. Avidan, F. Pozzi, M. Sorel, R. Morandotti, D. N. Christodoulides, and Y. Silberberg, Phys. Rev. Lett. **100**, 013906 (2008).
12. A. Szameit, Y. Kartashov, P. Zeil, F. Dreisow, M. Heinrich, S. Nolte, A. Tünnermann, V. Vysloukh, and L. Torner, Opt. Lett. **35**, 1172 (2010).
13. L. Martin, G. Di Giuseppe, A. Perez-Leija, R. Keil, F. Dreisow, M. Heinrich, S. Nolte, A. Szameit, A. F. Abouraddy, D. N. Christodoulides, and B. E. A. Saleh, Opt. Exp. **19**, 13636 (2011).
14. E. Castano and Y. C. Lee, Chin. J. Phys. **23**, 245 (1985).
15. A. Szameit, Y. V. Kartashov, F. Dreisow, M. Heinrich, T. Pertsch, S. Nolte, A. Tünnermann, V. A. Vysloukh, F. Lederer, and L. Torner, Phys. Rev. Lett. **102**, 063902 (2009).
16. D. M. Jovic, M. R. Belic, and C. Denz, Phys. Rev. A **84**, 043811 (2011).
17. A. Szameit and S. Nolte, J. Phys. B **43**, 163001 (2010).

**References with titles**

1. W. E. Buhro and V. L. Colvin, "Semiconductor nanocrystals: Shape matters," Nature Mater. **2**, 138 (2003).
2. L. Esaki and R. Tsu, "Superlattice and Negative Differential Conductivity in Semiconductors," IBM J. Res. Dev. **14**, 61 (1970).
3. H. Rodriguez-Coppola, "The dielectric response function of systems with reduced dimensionality," Microelectron. J. **33**, 379 (2002).
4. P. W. Anderson, "Absence of diffusion in certain random lattices," Phys. Rev. **109**, 1492–1505 (1958).
5. D. S. Wiersma, P. Bartolini, A. Lagendijk, and R. Righini, "Localization of light in a disordered medium," Nature **390**, 671 (1997).
6. J. Billy, V. Josse, Z. Zuo, A. Bernard, B. Hambrecht, P. Lugan, D. Clement, L. Sanchez-Palencia, P. Bouyer, and A. Aspect, "Direct observation of Anderson localization of matter waves in a controlled disorder," Nature **453**, 891 (2008).
7. G. Roati, C. D'Errico, L. Fallani, M. Fattori, C. Fort, M. Zaccanti, G. Modugno, M. Modugno, and M. In guscio, "Anderson localization of a non-interacting Bose-Einstein condensate," Nature **453**, 895 (2008).
8. H. Hui, A. Strybulevych, J. H. Page, S. E. Skipetrov, and B. A. van Tiggelen, "Localization of ultrasound in a three-dimensional elastic network," Nature Phys. **4**, 945 (2008).
9. J. Chabe, G. Lemarie, B. Gremaud, D. Delande, P. Szriftgiser, and J. C. Garreau, "Experimental Observation of the Anderson Metal-Insulator Transition with Atomic Matter Waves," Phys. Rev. Lett. **101**, 255702 (2008).
10. T. Schwartz, G. Bartal, S. Fishman and M. Segev, "Transport and Anderson localization in disordered two-dimensional photonic lattices," Nature **466**, 52 (2007).
11. Y. Lahini, A. Avidan, F. Pozzi, M. Sorel, R. Morandotti, D. N. Christodoulides, and Y. Silberberg, "Anderson Localization and Nonlinearity in One-Dimensional Disordered Photonic Lattices," Phys. Rev. Lett. **100**, 013906 (2008).
12. A. Szameit, Y. Kartashov, P. Zeil, F. Dreisow, M. Heinrich, S. Nolte, A. Tünnermann, V. Vysloukh, and L. Torner, "Wave localization at the boundary of disordered photonic lattices," Opt. Lett. **35**, 1172 (2010).
13. L. Martin, G. Di Giuseppe, A. Perez-Leija, R. Keil, F. Dreisow, M. Heinrich, S. Nolte, A. Szameit, A. F. Abouraddy, D. N. Christodoulides, and B. E. A. Saleh, "Anderson localization in optical waveguide arrays with off-diagonal coupling disorder," Opt. Exp. **19**, 13636 (2011).
14. E. Castano and Y. C. Lee, "Effect of Finite Width on Anderson Localization in a Strip," Chin. J. Phys. **23**, 245 (1985).
15. A. Szameit, Y. V. Kartashov, F. Dreisow, M. Heinrich, T. Pertsch, S. Nolte, A. Tünnermann, V. A. Vysloukh, F. Lederer, and L. Torner, "Soliton excitation in waveguide arrays with an effective intermediate dimensionality," Phys. Rev. Lett. **102**, 063902 (2009).
16. D. M. Jovic, M. R. Belic, and C. Denz, "Transverse localization of light in nonlinear photonic lattices with dimensionality crossover," Phys. Rev. A **84**, 043811 (2011).
17. A. Szameit and S. Nolte, "Discrete optics in femtosecond-laser-written photonic structures," J. Phys. B **43**, 163001 (2010).